\documentclass[reqno,11pt]{amsart}
\usepackage[utf8]{inputenc}
\usepackage{enumitem}
\usepackage{graphicx}
\usepackage{amscd}
\usepackage{slashed}
\usepackage{amssymb}
\usepackage{mathtools} 
\usepackage{pstricks}
\usepackage{setspace}
\usepackage{bbm}
\usepackage[framemethod=tikz]{mdframed}
\usepackage{csquotes}
\usepackage{framed}
\usepackage[mathscr]{eucal}
\numberwithin{equation}{section}
\allowdisplaybreaks[1]

\renewenvironment{leftbar}[1][\hsize]
{%
	\MakeFramed{\hsize#1\advance\hsize-\width\FrameRestore}%
}
{\endMakeFramed}

\title[{An Elementary Introduction to Causal Fermion Systems}]{Causal Fermion Systems: \\
An Elementary Introduction to \\ Physical Ideas and Mathematical Concepts}

\author[F.\ Finster]{Felix Finster}
\author[M.\ Jokel]{Maximilian Jokel \\ \\ August 2019}
\address{Fakult\"at f\"ur Mathematik \\ Universit\"at Regensburg \\ D-93040 Regensburg \\ Germany}
\email{finster@ur.de, maximilian.jokel@ur.de}

\newtheorem{Def}{Definition}[section]

\newcommand{\Thanks}{\vspace*{.5em} \noindent \thanks}
\newcommand{\beq}{\begin{equation}}
\newcommand{\eeq}{\end{equation}}
\newcommand{\Proof}{\begin{proof}}
\newcommand{\QED}{\end{proof} \noindent}

\newcommand{\la}{\langle}
\newcommand{\ra}{\rangle}

\newcommand{\Sl}{\mathopen{\prec}}
\newcommand{\Sr}{\mathclose{\succ}}
\newcommand{\C}{\mathbb{C}}
\newcommand{\R}{\mathbb{R}}
\newcommand{\1}{\mbox{\rm 1 \hspace{-1.05 em} 1}}

\newcommand{\N}{\mathbb{N}}
\renewcommand{\H}{\mathscr{H}}

\newcommand{\F}{{\mathscr{F}}}
\newcommand{\Dir}{{\mathcal{D}}}

\renewcommand{\L}{{\mathcal{L}}}
\newcommand{\Sact}{{\mathcal{S}}}
\newcommand{\T}{{\mathcal{T}}}

\newcommand{\Lin}{\text{\rm{L}}}

\newcommand{\scrM}{\myscr M}
\newcommand{\scrN}{\myscr N}

\newcommand{\bitem}{\begin{itemize}[leftmargin=2.5em]}
\newcommand{\eitem}{\end{itemize}}
\newcommand{\itemD}{\item[{\raisebox{0.125em}{\tiny $\blacktriangleright$}}]}

\DeclareFontFamily{OT1}{rsfso}{}
\DeclareFontShape{OT1}{rsfso}{m}{n}{ <-7> rsfso5 <7-10> rsfso7 <10-> rsfso10}{}
\DeclareMathAlphabet{\myscr}{OT1}{rsfso}{m}{n}

\newcommand{\Cb}{\textcolor{blue!80!black}}

\newcommand{\p}{\mathfrak{p}}
\newcommand{\q}{\mathfrak{q}}
\newcommand{\U}{\text{\rm{U}}}

\DeclareMathOperator{\Tr}{Tr}
\DeclareMathOperator{\tr}{tr}

\DeclareMathOperator{\supp}{supp}

\begin{document}

\maketitle

\begin{abstract}
We give an elementary introduction to the theory of causal fermion systems,
with a focus on the underlying physical ideas and the conceptual and mathematical foundations.
\end{abstract} 

\tableofcontents

%\newpage

\section[Unifying Quantum Field Theory and General Relativity]{The Challenge: Unifying Quantum Field Theory and General Relativity}
One of the biggest problems of present-day theoretical physics is the incompatibility of Quantum Field Theory and General Relativity. While the standard model of elementary particle physics provides a
quantum field theoretical description of matter together with its electromagnetic, weak and strong interactions 
down to atomic and subatomic scales,
General Relativity applies to gravitational phenomena on astrophysical or cosmological scales. Just as the standard model of elementary particle physics is well-confirmed by high-precision measurements, also the theoretical predictions of General Relativity agree with the experimental results to high accuracy.
Nevertheless, when combining Quantum Field Theory and General Relativity on very small length scales,
these theories become mathematically inconsistent, making physical predictions impossible.
%However, when describing processes and phenomena which both involve high energies and take place on mesoscopic scales, thus falling within the scope of both theories, mathematical inconsistencies make physical predictions impossible. %Typical examples are the ultraviolet divergences occurring in the calculation of Feynman diagrams or the curvature singularity at the center of a black hole. Both phenomena have in common that the reason of the occurrence of divergences lies in our not only experimental but in principle incomplete understanding of the microstructure of spacetime.

The fact that combining Quantum Field Theory with General Relativity leads to 
inconsistencies,
%meaningless predictions in the form of divergent expressions for physical quantities
although each theory by itself provides excellent theoretical predictions, allows for different possible
conclusions: While a convinced elementary particle physicist will refer to the overwhelming triumph of Quantum Field Theory and postulate the existence of a gravitational exchange particle, namely the graviton, thus forcing General Relativity into the setting of the standard model of elementary particle physics, a dedicated relativist, on the other hand, will question the mathematical formalism of Quantum Field Theory and instead refer to the aesthetics and mathematical
clarity of the differential geometric approach to General Relativity. Undecided physicists, who are convinced of the concepts of both Quantum Field Theory and General Relativity, may argue that, instead of
incorporating one theory in the other, one should try to find a new theory which reproduces both
Quantum Field Theory and General Relativity in suitable limiting cases.
Physicists skeptical of both theories will bring into play alternative approaches such as string theory or the theory of loop quantum gravity which are based on completely new assumptions.

Due to the lack of experimental evidence, most alternative approaches are
mainly based on personal preferences and paradigms. They involve ad-hoc assumptions which are
often detached from the well-established physical principles which were developed based on physical experiments.
Since there are many ways to introduce new assumptions ad hoc, it is questionable whether
these approaches will turn out to be successful.
Therefore, we prefer to proceed differently as follows:
We begin with a detailed and honest review of the concepts and principles which form the basis
of Quantum Field Theory and General Relativity.
Afterward, we select those principles which we consider to be essential
(clearly, this is a subjective choice).
Then we combine these principles in a novel mathematical setting, referred to as causal fermion systems.
Working exclusively with the objects in this setting, we postulate new physical equations by formulating
the so-called causal action principle. The causal action principle gives rise to additional objects and
structures in space-time together with equations describing their dynamics.
In this way, we obtain a new physical theory with predictive power.

\section{Overview of Concepts and Mathematical Structures in Theoretical Physics}
Following the above outline, this section is devoted to a review of the concepts and ideas, common beliefs as well as selected mathematical structures and objects used in contemporary theoretical physics. To sharpen the view for the few really fundamental principles underlying our present understanding and mathematical description of nature, we have decided to take a bird's-eye perspective rather than a high-resolution examination of sophisticated mathematical constructions.

\subsection{The Fabric of Spacetime}
Before Einstein's Special Theory of Relativity, physicists thought of space as being the 
geometric background in which physical processes take place while time evolves.
%In much the same way as a theatre provides a solid stage for various performances with different sceneries in the course of time, also space provides the invariable basis for an infinite variety of processes.
With this concept of space and time in mind, nobody could imagine that space itself
might change while time evolves or -- even more -- that space-time as a whole participates in the
physical interactions. After more than one hundred years of studying Einstein's Theory of Relativity, however, our understanding of space and time has changed completely.
Nowadays, we are used to the fact that space-time and its matter content 
cannot be considered independently, but rather form an inseparable unity
interwoven by mutual interactions. This unity is sometimes referred to as the \textit{fabric of spacetime}.

We now review the necessary concepts to capture and cast this intuitive notion in a formal mathematical
framework as provided by differential geometry.
In order to make this paper accessible to a broad readership, we also recall basic definitions
which are clearly familiar to mathematicians.

%In analogy to the human mind, which can not be considered detached from the brain in the sense of an operating system, also processes in nature cannot be considered detached from spacetime. Just as the human mind is some process within the brain tissue,
%physical processes are activities in the universe tissue which we refer to as the \textit{fabric of spacetime}.
%With this picture in mind, we want to review the necessary concepts to capture and cast this intuitive notion in a formal %mathematical framework as provided by differential geometry.

%Einstein's General Relativity is formulated in the language of differential geometry.
%As realized by Einstein, the appropriate mathematical language
%to describe spacetime in General Relativity is provided by differential geometry.
%We now recall a few basic concepts.

\subsubsection{Topological Manifolds as Models of Spacetime} At the most basic level, namely without considering any additional structures, spacetime is nothing but a set of points which locally -- that is within the tiny snippet of the universe which is accessible to our everyday experience -- looks like the familiar, three-dimensional Euclidean space. Including time as a fourth dimension naturally leads to the idea to model the fabric of spacetime as a four-dimensional topological manifold. 
% In physical parlance, the elements of the set $\scrM$ are events in the sense of the Special Theory of Relativity.

\begin{Def}[Topological Manifold]
\begin{leftbar}
		\normalfont A \textit{topological manifold of dimension~$d$} is a second-countable, topological Hausdorff space	$(\scrM,\mathcal{O})$ which at every point $p\in\scrM$ has a neighborhood which is homeomorphic to an open subset of $\R^d$.
\end{leftbar}
\end{Def}
\vspace*{2.0mm}
\noindent
Here~$\mathscr{O}$ denotes the family of all open subsets of~$\scrM$.
The reason why we do not consider a completely structureless set rather than the tuple $(\scrM,\mathcal{O})$ consisting of a set equipped with a topology, is needed in order to have a notion of continuity.

\subsubsection{Establishing Smooth Structures in Spacetime}
By modelling spacetime as a four-dimensional topological manifold, we have already implemented some of 
our knowledge about the general structure of our Universe. In order to describe smooth functions in spacetime and to be able to do calculus, one important ingredient is still missing and calls for the following definition:

\begin{Def}[Smooth Compatibility of Coordinate Charts]
\begin{leftbar}
		\normalfont Let $(\scrM,\mathcal{O})$ be an $d$-dimensional topological manifold together with two coordinate charts $(U,\varphi)$ and $(V,\psi)$ such that the open sets $U,V\subseteq\R^n$ satisfy $U\cap V\neq\emptyset$.\\[-5.0mm]
		\begin{center}
		\rule{\textwidth}{0.8pt}\\
		\end{center}
		\vspace*{1.0mm}
		The composition of the coordinate functions given by
			\begin{equation*}
				\psi\circ\varphi^{-1}\!:\,\varphi(U\cap V)\to\psi(U\cap V)
			\end{equation*}
		is called \textit{transition map from $\varphi$ to $\psi$}. Two coordinate charts $(U,\varphi)$ and $(V,\psi)$ are \textit{smoothly compatible} if the transition map $\psi\circ\varphi^{-1}$ is a diffeomorphism.
\end{leftbar}
\end{Def}
\vspace*{2.0mm}
\noindent
The definition of smoothly compatible coordinate charts allows us to introduce the notion of smooth atlases which in turn prepares the ground for defining smoothness of functions on manifolds.

\begin{Def}[Smooth Atlas]
\begin{leftbar}
		\normalfont Let $\{(U_i,\varphi_i)\}_{i\in I}$ with $I\subseteq\N$ be a family of charts of a topological manifold $(\scrM,\mathcal{O})$ with open sets $U_i\subseteq R^n$.\\[-5.0mm]
		\begin{center}
			\rule{\textwidth}{0.8pt}\\
		\end{center}
		\vspace*{1.0mm}
		The family $\{(U_i,\varphi_i)\}_{i\in I}$ of charts is called \textit{atlas}, if the open sets $U_i$ cover $\scrM$. If in addition any two charts in the atlas are smoothly compatible, the atlas is referred to as \textit{smooth atlas}.
\end{leftbar}
\end{Def}
\vspace*{2.0mm}
\noindent
A topological manifold equipped with a smooth atlas~${\mathscr{A}}$ is referred to as a {\em{smooth manifold}}.
We can now specify what we mean by smoothness of functions on a manifold.

\begin{Def}[Smooth Functions on Manifolds]
\begin{leftbar}
		\normalfont Let $(\scrM,\mathscr{A})$ be a smooth manifold.\\[-5.0mm]
		\begin{center}
			\rule{\textwidth}{0.8pt}\\
		\end{center}
		\vspace*{1.0mm}
		A function $f\!:\,\scrM\to\R$ on the manifold is called \textit{smooth} if for every chart $(U,\varphi)
		\in {\mathscr{A}}$ the function $f\circ\varphi^{-1}$ is smooth in the sense of functions being defined on open subsets of $\R^d$.
\end{leftbar}
\end{Def}

\subsubsection{Encoding the Lorentzian Geometry of Spacetime}
From our everyday life within a small snippet of the universe, we are used to the properties of three-dimensional Euclidean space, especially its vector space character. In order not to loose these familiar and useful properties when modelling spacetime as a differentiable manifold, one introduces a vector space structure at every single point of the manifold. In order to avoid the idea that spacetime is embedded in some higher-dimensional ambient space, we must work with an intrinsic characterization which only makes use of the already defined
concepts of coordinate charts and smooth functions.
	\begin{Def}[Derivations and Tangent Space]
\begin{leftbar}
		\normalfont Let $(\scrM,\mathscr{A})$ be an $d$-dimensional smooth manifold and $p$ an element of $\scrM$.\\[-5.0mm]
		\begin{center}
			\rule{\textwidth}{0.8pt}\\
		\end{center}
		\vspace*{1.0mm}
		A linear map $X_p\!:C^\infty(\scrM,\R)\to\R$ is called \textit{derivation at $p\in\scrM$} if it satisfies the Leibniz product rule
			\begin{equation*}
				\forall f,g\in C^\infty(\scrM,\R)\!:\,X_p(fg)=f(p)X_p(g)+g(p)X_p(f)
			\end{equation*}
		The set of all derivations at $p\in\scrM$ forms a vector space under the operations
			\begin{align*}
				(X+Y)_p(f)&=X_p(f)+Y_p(f)\\
				(\alpha X)_p(f)&=\alpha X_p(f)
			\end{align*}
		which is referred to as the \textit{tangent space $T_p\scrM$ at $p\in\scrM$}.
\end{leftbar}
\end{Def} \nopagebreak
\vspace*{2.0mm}
\noindent
It can be shown that the tangent space is a $d$-dimensional real vector space.
In order to better understand the similarities between the differential geometric formulation of Einstein's General Theory of Relativity and the theory of causal fermion systems in the further course of this article, we shall introduce the bundle formulation.
\begin{leftbar}
	\begin{Def}[Tangent Bundle and Vector Fields] \label{deftangentbundle}
		\normalfont Let $(\scrM,\mathscr{A})$ be an $d$-dimensional smooth manifold with tangent spaces $T_p\scrM$ at all points $p\in\scrM$.\\[-5.0mm]
		\begin{center}
			\rule{\textwidth}{0.8pt}\\
		\end{center}
		\vspace*{1.0mm}
		\noindent
		The \textit{tangent bundle $T\scrM$} is defined as the disjoint union of the tangent spaces $T_p\scrM$ at all points $p\in\scrM$
			\begin{equation*}
				T\scrM:=\bigcup\limits_{p\in\scrM}\{p\}\times T_p\scrM
			\end{equation*}
		(endowed with the coarsest topology which makes the bundle charts continuous).
		A continuous function $X\in C^0(\scrM,T\scrM)$ is called \textit{vector field} if it satisfies the condition
			\begin{equation*}
				\forall p\in\scrM\!:\,X(p):=X_p\in T_p\scrM
			\end{equation*}
	\end{Def}
\end{leftbar}
\vspace*{2.0mm}
\noindent
Having defined tangent spaces, we are ready to add our knowledge about the geometric structure of spacetime to our model. In the familiar Euclidean geometry, the geometry is retrieved by computing lengths and angles between
vectors. These quantities are encoded in a scalar product, being a positive definite bilinear form
\[ g_p\!:\,T_p\scrM\times T_p\scrM\to\R \:. \]
In Special Relativity, the geometry is described again by a bilinear form, which however
is no longer positive definite, but instead has signature~$(1,3)$:
\begin{Def}[Lorentzian Manifold]
\begin{leftbar}
		\normalfont Let $(\scrM,\mathscr{A})$ be a $d$-dimensional smooth manifold with tangent bundle $T\scrM$.\\[-5.0mm]
		\begin{center}
			\rule{\textwidth}{0.8pt}\\
		\end{center}
		\vspace*{1.0mm}
		\noindent
		A function $g\!:\,T\scrM\times T\scrM\to\R$ is called \textit{Lorentzian metric} if the restriction 
			\begin{equation*}
				g_p\!:\,T_p\scrM\times T_p\scrM\to\R
			\end{equation*}
		is a bilinear, symmetric and smooth mapping
			\begin{equation*}
				g(X,Y)\!:\,\scrM\to\R\qquad\qquad p\mapsto\big[g(X,Y)\big](p):=g_p(X_p,Y_p)
			\end{equation*}
		of signature~$(1,3)$.
		A smooth manifold $(\scrM,\mathscr{A})$ equipped with a Lorentzian metric is referred to as \textit{Lorentzian manifold $(\scrM,g)$}.
\end{leftbar}
\end{Def}
%\vspace*{2.0mm}
%\noindent
%Among the pseudo-Riemannian manifolds, those whose metric tensor have signature $(1,d-1)$ are referred to as \textit{Lorentzian manifolds}. 
The Lorentzian signature implies that the inner product~$g_p(\xi, \xi)$
of a tangent vector~$\xi \in T_p\scrM$ with itself can be positive or negative. This gives rise to the following
notion of {\em{causality}}. A tangent vector $\xi \in T_p\scrM$ is said to be
\beq \label{causalMink}
\left\{ \begin{array}{cll}
{\mbox{\em{timelike}}} &\quad& {\text{if $g_p(\xi, \xi) >0$}} \\
{\mbox{\em{spacelike}}} && {\text{if $g_p(\xi, \xi) <0$}} \\
{\mbox{\em{lightlike}}} && {\text{if $g_p(\xi, \xi) =0$}}\:. \end{array}  \right.
\eeq
Lightlike vectors are also referred to as null vectors, and the term {\em non-spacelike} refers to timelike or lightlike vectors.
The spacetime trajectory of a moving object is described by a curve
$\gamma(\tau)$ in~$\scrM$ (with $\tau$ an arbitrary parameter).
We say that the spacetime curve $\gamma$ is timelike if the tangent vector~$\dot{\gamma}(\tau)$
is everywhere timelike. Spacelike, null and
non-spacelike curves are defined analogously. Then 
the usual statement of causality that nothing
can travel faster than the speed of light can be formulated as follows: 
\medskip
\begin{quote}
{\em{Causality}}: \quad \parbox[t]{8.5cm}{
Information can be transmitted only along
non-space\-like curves.}
\end{quote}\medskip

\subsection{The Einstein Field Equations}
After these preparatory considerations, we are now ready to formulate and investigate the significance of the Einstein field equations which are at the heart of General Relativity. They take the form
\[ %\beq \label{einstein}
\text{Ric} - \frac{1}{2}\: R\: g + \Lambda\, g = 8 \pi \kappa\, T \:, \]
where~$\text{Ric}$ is the Ricci tensor, $R$ is scalar curvature, $\Lambda$ is the cosmological constant,
$\kappa$ is the gravitational coupling constant, and~$T$ is the energy-momentum tensor.
These equations can be derived from an action principle. More precisely, metrics which
solve the Einstein equations are critical points of the Einstein-Hilbert action
\beq \label{EH}
\Sact^\text{EH} = \int_\scrM \Big( \frac{1}{16 \pi \kappa}\:\big( R - 2 \Lambda \big) +  \L_{\text{matter}} \Big)\: d\mu_\scrM(x)\:.
\eeq
The Einstein equations relate the curvature of spacetime (on the left side) to the matter distribution
described by the energy-momentum tensor (on the right side).
Combining the field equations with the equations of motion for the matter fields
(like the geodesic equation, the Dirac equation, etc.), one gets a coupled system of partial
differential equations. This coupled system can be understood
in simple terms by the popular phrase
that matter tells spacetime how to curve, and spacetime tells matter how to move.

Taking up the comparison between the brain-mind-relationship and the interplay of spacetime and physical processes therein, the Einstein field equations characterize this interrelation. 
In a similar way as our thinking
shapes the brain structures which in turn have an influence on our thoughts, also the objects existing and processes happening in spacetime deform spacetime, which has a back effect on physical processes. Einstein's revolutionary insight that spacetime together with its matter and energy content form an inseparable unity, is one of the cornerstones which the theory of causal fermion systems is built on.

\subsection{Quantum Theory in a Classical Spacetime}
The second groundbreaking discovery in the twentieth century besides General Relativity was Quantum Theory.
The insight that certain physical quantities take discrete rather than continuous values
revolutionized our understanding of Nature. This discovery triggered the development of Quantum Mechanics which is the appropriate framework to study the quantum behaviour of a single particle or a constant finite number of particles. Although the framework of Quantum Mechanics is appropriate to describe even arbitrarily large quantum
systems of a fixed number of particles, it is in principle incapable to formalize processes involving a varying number of quantum particles. This limitation is overcome in {\em{Quantum Field Theory}}, where a quantum state
can be a superposition of components involving a varying and arbitrarily large number of particles.

Relativistic Quantum Field Theory is usually formulated in {\em{Minkowski space}}, disregarding the
gravitational field (see for example~\cite{bjorken, bjorken2, peskin+schroeder}).
The fact that in Quantum Field Theory one deals with arbitrarily large number of particles which
can have arbitrarily large momenta can be understood as the reason why
divergences occur in the perturbative description.
The {\em{renormalization program}} provides a systematic
computational procedure for dealing with these divergences.
The such-renormalized Quantum Field Theory makes excellent physical predictions which have
been confirmed experimentally to high precision.
Nevertheless, it is often criticized that the renormalization program lacks a foundational justification.
Also, it is not quite satisfying that the theory is well-defined only to every order in perturbation theory.
But the perturbation series does not need to converge. Also, it is not clear whether
there exists a mathematically meaningful non-perturbative formulation of
Quantum Field Theory.

Most methods of Quantum Field Theory also apply to Quantum Field Theory in a
fixed {\em{curved spacetime}} (see for example~\cite{brunetti+fredenhagenQFT}
and the references therein).
In other words, one considers Quantum Fields in
the background of a {\em{classical}} gravitational field
without taking into account the backreaction of the quantum fields
to classical gravity.

%
%
%makes mathematical sense non-perturbatively.
%
%But the convergence of the perturbation series is unknown.
%
%depending on the individual conceptional standards,
%
%the renormalization procedure still 
%However, this procedure lacks a foundational justification.
%Moreover, this method works only for a class of theories called renormalizable.
%
%In this picture, the physical system without any particles is described mathematically by
%the vacuum state of the respective quantum field, while states with particles are interpreted as
%excitations of the vacuum, obtained
%by acting on the vacuum state with the field operator of the quantum field.
%
%
%From a conceptual point of view,
%particles by themselves are no longer considered as fundamental quantities.
%
%
%

%Instead, they are considered merely as a manifestation of the underlying quantum field,
%which in this way acquires physical significance.

\subsection{Incompatibility of General Relativity and Quantum Field Theory} \label{secinconsist}
Quantum field theory in a classical spacetime has the shortcoming that
classical and quantum objects coexist in a way which is conceptually not fully convincing.
It would be desirable to describe all the objects on the same footing
by ``unifying'' the theories. However, there is no consensus on how this
``unification'' should be carried out,
or even on what ``unification'' should mean.
Nevertheless, most physicists agree that serious problems arise,
no matter which approach for ``unification'' is taken.
In order not to take sides, we here merely list some of the most common
arguments pointing towards the difficulties:
\bitem
\itemD The simplest method is to start from the Heisenberg Uncertainty Principle
$\Delta p \, \Delta x \geq \frac{\hbar}{2}$, which states that position and
momentum of a point particle in quantum mechanics can be determined
simultaneously only up to a fundamental uncertainty given by Planck's
constant~$\hbar$. In Quantum Field Theory, similar uncertainty relations hold
for the field operators and the associated canonical momentum operators. In particular,
acting with the local field operator~$\phi(x)$ on the vacuum state,
the quantum state is localized in space, meaning that there is
a large momentum uncertainty. This also gives rise to a large
uncertainty in the corresponding energy. Intuitively speaking, we thus obtain
large ``fluctuations'' of energy in a small spatial region.
In General Relativity, on the other hand, high energy densities lead to
the formation of black holes. Therefore, combining the principles
of General Relativity and Quantum Field Theory in a naive way
leads to the formation of microscopic black holes, implying that the
concept of a spacetime being ``locally Minkowski space'' breaks down.
The relevant length scale for such effects is the 
{\em{Planck length}}~$\ell_P \approx 1.6 \times 10^{-35}\ \mathrm{m}$.
\itemD  The renormalization program only applies to a class of theories called renormalizable.
It turns out that applying the canonical quantization methods
to Einstein's gravity, the resulting theory is {\em{not}} renormalizable.
This shows that quantizing gravity with the present methods of perturbative Quantum Field Theory 
is not a fully convincing concept.
\itemD It is sometimes argued that the problem of ``unification''
is rooted in shortcomings of present Quantum Field Theory.
Indeed, the ultraviolet divergences of Quantum Field Theory
suggest that the structure of spacetime should be modified for very small
distances. A natural length scale for such modifications is given by the
Planck length. In this way, the problem of the ultraviolet divergences seems to
be intimately linked to gravity.
Therefore, in order to resolve these problems, one should modify the
structure of spacetime on the Planck scale.
\eitem

\subsection{A Step Back: Quantum Mechanics in Curved Spacetime} \label{secQMcurved}
In order to avoid the just-described problems which arise when ``unifying''
General Relativity with Quantum Field Theory, it is a good idea to
take a step back and
return to the familiar and well-understood grounds of one-particle {\em{quantum mechanics}}.
Indeed, formulating quantum mechanics in curved spacetime does not lead to any
conceptual or technical problems. We now review a few basic concepts,
which will also be our starting point for the constructions leading to causal fermion systems.

\subsubsection{The Dirac Equation in Minkowski Space}
In non-relativistic quantum mechanics, a particle is described
by its Schr\"odinger wave-function~$\psi(t,\vec{x})$.
It has a probabilistic interpretation, meaning that its
absolute square~$|\psi(t,\vec{x})|^2$ is the probability density for the particle to be at the position~$\vec{x} \in \R^3$.
A relativistic generalization of the Schr\"odinger equation is the {\em{Dirac equation}}.
In this case, the wave function~$\psi$ has four complex components, which describe the spin of the particle.
In flat Minkowski space, the Dirac equation takes the form
\beq
    \Big( i \gamma^k \frac{\partial}{\partial x^k} - m \Big)
    \psi(x) = 0 \;, \label{Dirac1}
\eeq
where~$x=(t,\vec{x}) \in \scrM$ is a point of Minkowski space, $m$ is the rest mass,
and the so-called Dirac matrices~$\gamma^k$ are $4 \times 4$-matrices which are 
related to the Lorentzian metric by the {\em{anti-commutation relations}}
\[ %\beq
2\: g^{jk}\:\1 = \{ \gamma^j,\: \gamma^k\} \;\equiv\; \gamma^j \gamma^k +
\gamma^k \gamma^j \:. %\label{f:0b}
\]
The Dirac spinors at every spacetime point are
endowed with an indefinite inner product of signature $(2,2)$, which we call
{\em{spin scalar product}} and denote by~$\Sl \psi | \phi \Sr(x)$.
To every solution~$\psi$ of the Dirac equation
we can associate a vector field $J$ by
\[ %\beq \label{dc}
J^k = \Sl \psi \,|\, \gamma^k\: \psi \Sr \;, \]
referred to as the {\em{Dirac current}}.
For solutions of the Dirac equation, this vector field is divergence-free.
This is referred to as {\em{current conservation}}.

Current conservation is closely related to the probabilistic interpretation of
the Dirac wave function. Indeed, as a consequence of current conservation,
for a solution~$\psi$ of the Dirac equation, the spatial integral
\[ %\beq
( \psi | \psi ) := 2 \pi \int_{\R^3} \Sl \psi \,|\, \gamma^0 \psi
\Sr(t, \vec{x}) \:d^3x %\label{printMink}
\]
is time independent. Normalizing the value of this integral to one, its integrand
gives the probability density of the particle to be at position~$\vec{x}$.

\subsubsection{The Dirac Equation in Curved Spacetime}
In curved spacetime, the Dirac equation is described most conveniently
using vector bundles. Similar to the tangent bundle in Definition~\ref{deftangentbundle},
the spinor bundle is obtained by ``attaching'' a vector space~$S_p\scrM$ to every spacetime point,
\[ S\scrM = \bigcup\limits_{p\in\scrM}\{p\}\times S_p\scrM \:. \]
But now, the vector space~$S_p\scrM$, the so-called {\em{spinor space}}, is a four-dimensional
complex vector space. This vector space is endowed with an indefinite inner product of signature~$(2,2)$
which, just as in Minkowski space, we refer to as the {\em{spin scalar product}} and denote by
\[ \Sl .|. \Sr_p \::\: S_p\scrM \times S_p\scrM \rightarrow \C \:. \]
At each spacetime point~$p$, the Dirac wave function~$\psi$ takes
a value in the corresponding spinor space~$S_p\scrM$.
The Dirac operator~$\Dir$ takes the form
\[ \Dir := i \gamma^j \nabla_j \:, \]
where~$\nabla_j$ is a connection on the spinor bundle, and the Dirac matrices are
related to the Lorentzian metric again by the anti-commutation relations
\[ \{ \gamma^j(p),\: \gamma^k(p) \} = 2\: g^{jk}(p)\:\1_{S_p\scrM} \:. \]
The Dirac equation in curved spacetime reads
\[ %\beq \label{Dirac}
(\Dir - m) \,\psi = 0 \:. \]
On solutions of the Dirac equation, one has the scalar product
\beq \label{print}
(\psi | \phi)_m := \int_\scrN \Sl \psi \,|\, \nu^j \gamma_j\, \phi \Sr_p\: d\mu_\scrN(p) \:,
\eeq
where~$\nu$ is the future-directed normal on the Cauchy surface~$\scrN$,
and~$d\mu_\scrN$ is the induced measure.
For mathematical completeness, we point out that 
we always assume that spacetime is globally hyperbolic, so that Cauchy
surfaces exist. Moreover, in order for the integral in~\eqref{print} to be finite,
we restrict attention to wave functions of spatially compact support
(i.e.\ to wave functions whose restriction to any Cauchy surface have compact support).

Due to current conservation,
the scalar product~\eqref{print} is independent of the choice of the Cauchy surface.
Choosing~$\psi=\phi$ as a unit vector,
the integrand of the above scalar product again has
the interpretation as the quantum mechanical probability density.

%
%(due to current conservation, the scalar product is
%in fact independent of the choice of~$\scrN$; for details see~\cite[Section~2]{finite}).
%Forming the completion gives the Hilbert space~$(\H_m, (.|.)_m)$.
%
%
%\begin{Def}[Spinors \& Spinor Bundles]
%\end{Def}
%
%\begin{Def}[Dirac Operator]
%\normalfont 
%\end{Def}
%
%

\section{Conceptual and Mathematical Foundations of Causal Fermion Systems}\label{sec:conceptfound}\noindent
The theory of causal fermion systems is a novel approach to fundamental physics which is built on our
conviction that, in order to resolve the incompatibility of General Relativity and Quantum Field Theory described above, one should \textit{modify the geometric structure of spacetime on microscopic scales}. Having already
surveyed our current way of modelling the fabric of spacetime and quantum wave functions therein, we now introduce the conceptual foundations of the theory of causal fermion systems.

\subsection{Guiding Principles of the Theory of Causal Fermion Systems}
Following Einstein's celebrated insight that \textit{\enquote{one cannot solve problems with the same level of thinking that created them,}} the theory of causal fermion systems does not try to force obviously incompatible concepts into an already existing setting, but instead provides a new mathematical framework which is inspired by carefully selected concepts from contemporary theoretical physics. The main guiding principles of the theory of causal fermion systems are the following ideas:
\bitem \setlength\itemsep{1mm}
%\begin{itemize}[leftmargin=2em]
%	\renewcommand{\labelitemi}{\raisebox{0.3mm}{\scalebox{0.6}{$\blacksquare$}}}
\itemD \textbf{Unified description of spacetime and the objects therein} \\
The General Theory of Relativity impressively demonstrates that seemingly disparate concepts such as the motion of matter and the metric tensor structure of spacetime are closely related and cannot be considered independent of each other. This surprising insight illustrates the geometric character, high degree of complexity and interconnectedness of the Universe. The~interdependence of matter distributions and the shape of spacetime which reacts on local changes as a whole, strongly suggests to take a \textit{unified point of view} when developing new physical theories. This fundamental conviction is implemented in the theory of causal fermion systems in that spacetime, together with all objects therein (such as particles, fields, etc.),
			are determined  dynamically as a whole by
minimizing the so-called \textit{causal action}. %\\[-1em]
\itemD \textbf{Equivalence principle} \\
In General Relativity, the equivalence principle is implemented
mathematically by working with geometric objects on a Lorentzian manifold.
In particular, the Einstein-Hilbert action is diffeomorphism invariant.
Allowing for a nontrivial microscopic structure, in the setting of causal fermion systems
spacetime does not necessarily need to be a smooth manifold.
Consequently, instead of diffeomorphisms, one must allow for more general
transformations of spacetime. The causal action is invariant under these more
general transformations, thereby generalizing the equivalence principle.
\itemD \textbf{Principle of causality} \\
The principle of causality plays a crucial role in our understanding of the structure of
physical interactions in spacetime. A guiding conception in the development of causal
fermion systems was that the causal structure of spacetime is not give a-priori, but that it is
determined dynamically when solving the physical equations.
For a better comparison, we recall that in General Relativity, the causal structure is
encoded in the Lorentzian metric (as explained after~\eqref{causalMink}).
Therefore, when varying the metric in the Einstein-Hilbert action~\eqref{EH}, also the causal structure
changes. Only after a critical point of the Einstein-Hilbert action has been found,
the corresponding metric determines the causal structure of spacetime.
Similarly, in the theory of causal fermion systems, the causal structure of spacetime is
determined only after a critical point of the causal action has been found.
%It is determined by the measure~$\rho$.
The principle of causality is implemented
in the form that points with spacelike separation are not related to each other in the
Euler-Lagrange equations corresponding to the causal action principle.
\itemD \textbf{Local gauge principle} \\
In classical electrodynamics, the local gauge principle
means the freedom~$A \rightarrow A + d\Lambda$ in changing the electromagnetic
potential~$A$ by the derivative of a scalar function~$\Lambda$.
This observation was the starting point for the development of gauge theories,
which have been highly successful in describing all the interactions in the
standard model. In Quantum Theory, local gauge transformations correspond to
generalized local phase transformations of the wave functions
\[ \psi(x) \rightarrow U(x)\: \psi(x) \:, \]
where~$U(x)$ is an isometry on the fibres of the spinor bundle.
The theory of causal fermion systems incorporates this principle
in that the causal action is invariant under such local transformations.
\itemD \textbf{Microscopic spacetime structure} \\
The ultraviolet divergences in Quantum Field Theory suggest that one should modify the
microscopic structure of spacetime. In order to include these microscopic features of spacetime, the theory of causal fermion systems does not assume physical spacetime to be continuous down to smallest scales,
but instead allows for a nontrivial, possibly discrete microstructure of spacetime.
\itemD \textbf{Fermionic building blocks} \\
From high energy physics we have a quite clear and consistent picture of the elementary building blocks of Nature which is formalized in the Standard Model of Particle Physics. In particular, we know that the fundamental matter particles are fermions while the forces are mediated by bosons.
Inspired by Dirac's concept that in the Minkowski vacuum a whole ``sea'' of fermions is present,
we consider the fermions as being more fundamental, whereas
bosons appear in our approach merely as a device to describe the interaction of the fermions.
\eitem
\noindent
Causal fermion systems evolved in the attempt to combine the above principles
in a simple and compact mathematical setting.

In the following sections we enlarge on each of the guiding principles and explain how they are formalized within the mathematical framework of the theory of causal fermion systems.

\subsection{Unified Description of Spacetime and the Objects Therein}
The basic conceptual idea underlying the theory of causal fermion systems consists in the belief that a successful unified theory must provide a unified description of the Universe in the sense that it does not treat spacetime separate from its matter and energy content. This central conception is inspired by the inseparable unity of spacetime and its matter-energy content as described by Einstein's field equations. In much the same way as the Einstein-Hilbert action singles out those metric tensors which are critical points of
the action and declares them to be the physically admissible choices, also the theory of causal fermion systems is based on a variational principle. Before we 
can formulate such a variational principle, we give the general definition of a causal fermion system
and explain it afterward.
\begin{Def}[Causal Fermion System]
\begin{leftbar}
		\normalfont A \textit{causal fermion system} of spin dimension~$n \in \N$ is a triple $(\H,\F,\rho)$ consisting of the following three mathematical structures:\\[-3.0mm]
		\begin{enumerate}
			\item $\H$ is a complex, separable Hilbert space $(\H,\la\cdot|\cdot\ra_\H)$.
			\item $\F$ is the subset of the Banach space $(\Lin(\H),\|\cdot\|)$ comprising
			 all self-adjoint operators on $\H$ of finite rank, which -- counting multiplicities -- have at most $n\in\N$ positive and at most $n\in\N$ negative eigenvalues.
			\item $\rho$ is a positive Borel measure $\rho: {\mathcal{B}} \to\R^+_0\cup\{\infty\}$ on~$\F$
			(where~${\mathcal{B}}$ is the $\sigma$-algebra generated by all open subsets of~$\F$).
			The measure~$\rho$ is referred to as the {\em{universal measure}}.
\end{enumerate}
\end{leftbar}
\end{Def}
\noindent
The connection of this definition to physics is not obvious.
%In precisely the same way as Lorentzian manifolds provide the arena for General Relativity without a priori having any physical meaning, also Causal Fermion Systems only set the scene for further mathematical considerations -- of course without carrying any physical~meaning.\\[3.0mm]
In order to convey a better, more intuitive understanding of this definition, let us have a detailed view on the individual ingredients. The structure of a complex Hilbert space $(\H,\la\cdot|\cdot\ra_\H)$ is a commonly used structure both in mathematics as well as in theoretical physics and should therefore need no further explanation. In contrast to this, the set~$\F$ as well as the measure~$\rho$ -- although familiar to mathematicians -- are 
not commonly used in theoretical physics. In order to make the theory of causal fermion systems easier accessible to interested physicists, we now explain these structures in greater detail.

\subsubsection{The Measure Space~$(\F, {\mathcal{B}}, \rho)$}
In contrast to what one might expect from the ordering in the above definition, the central
structure of a causal fermion system is not the Hilbert space $\H$ itself but rather the measure~$\rho$.
{\em{Measures}} appear in physics mainly as integration measures, like for example the measure~$d\mu = d^3x$
in the three-dimensional integral
\[ \int_{\R^3} f(x)\: d\mu(x) \]
(of a, say, continuous and compactly supported function~$f$).
In mathematics, the measure~$\mu$ is a mapping which to a subset~$\Omega \subset \R^3$ associates
its volume,
\[ \mu \::\: \Omega \mapsto \mu(\Omega) := \int_\Omega d^3x \:. \]
A central conclusion from measure theory is that it is mathematically not sensible to associate a measure
to every subset of~$\R^3$. Instead, one must distinguish a class of sufficiently ``nice'' subsets as being
{\em{measurable}}. The measurable sets form a {\em{$\sigma$-algebra}}, meaning that applying
any finite or countable number of set operations on measurable sets gives again a measurable set.
Here it suffices to always work with the {\em{Borel algebra}}, defined as the smallest $\sigma$-algebra
where all open sets are measurable.

A difference to usual integration measures is that the universal measure~$\rho$ is a {\em{measure on linear operators}}.
The starting point is the Banach space~$\Lin(\H)$ of all bounded linear operators on $\H$ together with the operator norm
\begin{equation} \label{onorm}
\|x\|:=\sup\big\{\|xu\|_\H\,\,\big|\,\,\|u\|_\H=1\big\} \:.
\end{equation}
The set~$\F$ is by definition a subset of~$\Lin(\H)$. We point out that~$\F$ is {\em{not}} a subspace
of~$\Lin{\H}$, because linear combinations of operators in~$\F$ will in general have rank
greater than~$2n$. But, being a closed subset of~$\Lin(\H)$, it is a {\em{complete metric space}} with the
distance function
\[ d \::\: \F \times \F \rightarrow \R^+_0 \:,\qquad d(x,y) := \|x-y\|\:. \]
We remark that~$\F$ is {\em{not}} a manifold, even if~$\H$ is finite-dimensional.
However, the subset of all operators of maximal rank
\[ \F^\text{reg} := \big\{ x \in \F \:\big|\: \text{$x$ has rank $2n$} \big\} \]
is dense in~$\F$ and indeed a smooth manifold of dimension
\[ \dim \F^{\text{reg}} = 4n\, \big( \dim \H - n \big) \:. \]
(for details see~\cite[Proposition~2.4.4]{intro}). Since in physical applications the dimension of~$\H$
is very large, $\F$ should be regarded as a subset of~$\Lin(\H)$ of very high dimension.

In order to define a measure $\rho$ on this set of operators, we must construct a $\sigma$-algebra.
The simplest choice (which also covers all cases of present physical interest) is to take the Borel algebra~${\mathcal{B}}$,
i..e\ the $\sigma$-algebra generated by all open subsets of~$\F$, with respect to the topology induced by the
operator norm~\eqref{onorm}.  The measure~$\rho$ makes it possible to integrate
a continuous (or Borel) function~$f : \F \rightarrow \R^+_0$ over~$\F$,
\[ \int_\F f(x)\: d\rho(x) \in [0, \infty] \:. \]
All familiar concepts from integration theory in~$\R^3$ also apply here. However, one should keep in mind
that we integrate over a set of operators of the Hilbert space
(in other words, the integration variable~$x$ is operator-valued).

\subsubsection{The Causal Action Principle} \label{seccap}
We are now in the position to define the causal Lagrangian and the causal action.
For any~$x, y \in \F$, the product~$x y$ is an operator
of rank at most~$2n$. We denote its non-trivial eigenvalues counting algebraic multiplicities
by~$\lambda^{xy}_1, \ldots, \lambda^{xy}_{2n} \in \C$
(more specifically,
denoting the rank of~$xy$ by~$k \leq 2n$, we choose~$\lambda^{xy}_1, \ldots, \lambda^{xy}_{k}$ as all
the non-zero eigenvalues and set~$\lambda^{xy}_{k+1}, \ldots, \lambda^{xy}_{2n}=0$).

\begin{Def}[Causal Lagrangian and Causal Action]
\begin{leftbar}
		\normalfont The \textit{causal Lagrangian} is a function defined as
			\begin{equation} \label{Ldef}
				\L:\F\times\F\to\R_0^+\qquad(x,y)\mapsto\L(x,y):=\frac{1}{4n}\sum\limits_{i,j=1}^{2n}\left(\big|\lambda_i^{xy}\big|-\big|\lambda_j^{xy}\big|\right)^2 \:.
			\end{equation}
		where $\big|\lambda_i^{xy}\big|$ denotes the absolute values of the eigenvalues $\lambda_i^{xy}$ of the operator product $xy$. \\[-5.0mm]
		\begin{center}
			\rule{\textwidth}{0.8pt}\\
		\end{center}
		\vspace*{1.0mm}
		\noindent		
		The \textit{causal action} is obtained by integrating the Lagrangian with respect to the universal measure,
			\[
				\Sact(\rho) :=\iint\limits_{\F\times\F}\mathcal{L}(x,y)\,\mathrm{d}\rho(x)\,\mathrm{d}\rho(y) \:. \]
\end{leftbar}
\end{Def} \noindent
Having defined the causal action, we can introduce the variational principle,
which is the core of the theory of causal fermion systems:
\begin{Def}[Causal Action Principle and Constraints]
\begin{leftbar}
		\normalfont The \textit{causal action principle} is to minimize~$\Sact$ by varying the universal measure
under the following constraints:
\begin{align}
\text{\em{volume constraint:}} && \rho(\F) = \text{const} \quad\;\; & \label{volconstraint} \\
\text{\em{trace constraint:}} && \int_\F \tr(x)\: d\rho(x) = \text{const}& \label{trconstraint} \\
\text{\em{boundedness constraint:}} && \T(\rho) := \iint_{\F \times \F} 
\bigg( \sum_{i=1}^{2n} |\lambda^{xy}_i| \bigg)^2\: d\rho(x)\, d\rho(y) &\leq C \:. \label{Tdef}
\end{align}
\end{leftbar}
\end{Def} \noindent
Here~$C$ is a given parameter (and~$\tr$ denotes the trace of a linear operator on~$\H$).
The constraints are needed in order to obtain a well-posed variational principle
without trivial minimizers.

Although the mathematical structure of the causal action principle can be understood
from general considerations (as will be outlined below), its detailed form is far from obvious.
It is the result of many computations and long considerations, which we cannot review here.
Instead, we note that the causal action was first proposed in~\cite[Section~3.5]{pfp}, based on
considerations outlined in~\cite[Sections~5.5 and~5.6]{pfp}.
The significance of the constraints became clear in the later mathematical analysis~\cite{continuum}.

Note that the universal measure~$\rho$ is the basic object in the theory of causal fermion systems.
It is a unified theory in the sense that all spacetime structures are encoded in and must be derived from this measure.
In other words, the measure~$\rho$ describes our universe as a whole.
This explains the name {\em{universal measure}}.

\subsection{The Equivalence Principle}
Let~$(\H, \F, \rho)$ be a causal fermion system of spin dimension~$n$ which minimizes the causal action,
respecting the constraints.
\begin{Def}[Spacetime] \label{defst} \normalfont 
\begin{leftbar}
{\em{Spacetime}}~$M$ is defined as the support of the universal measure,
		\normalfont
\[ M := \text{supp}\, \rho \subset \F \:. \]
\end{leftbar}
\end{Def} \noindent
Here the {\em{support}} of a measure is defined as the complement of the largest open set of measure zero, i.e.
\[ \supp \rho := \F \setminus \bigcup \big\{ \text{$\Omega \subset \F$ \,\big|\,
$\Omega$ is open and $\rho(\Omega)=0$} \big\} \:. \]
Thus the space-time points are symmetric linear operators on~$\H$.
On~$M$ we consider the topology induced by~$\F$ (generated by the $\sup$-norm~\eqref{onorm}
on~$\Lin(\H)$). Moreover, the universal measure~$\rho|_M$ restricted to~$M$ can be regarded as a volume
measure on space-time. This makes space-time to a {\em{topological measure space}}.

Let~$\Phi : M \rightarrow M$ be a {\em{homeomorphism}} of spacetime.
Given a Borel set~$\Omega \subset \F$, the preimage~$\Phi^{-1}(\Omega \cap M)$ is a Borel set of~$M$.
Therefore, we can define a new Borel measure~$\tilde{\rho}$ on~$\F$ by~$\tilde{\rho}(\Omega) := 
\rho(\Phi^{-1}(\Omega \cap M))$. This is the so-called {\em{push-forward measure}} denoted by
\[ \tilde{\rho} = \Phi_* \rho \:. \]
The causal action as well as all the constraints are invariant under the transformation
\[ M \rightarrow \Phi(M) \:,\qquad \rho \mapsto \tilde{\rho} \:. \]
This invariance generalizes the diffeomorphism invariance of General Relativity.
In this sense, the {\em{equivalence principle}} is implemented in the theory of causal fermion systems.

\subsection{Principle of Causality}
For any~$x, y \in M$, the product~$x y$ is an operator
of rank at most~$2n$. Exactly as defined at the beginning of Section~\ref{seccap},
we denote its non-trivial eigenvalues (counting algebraic multiplicities)
by~$\lambda^{xy}_1, \ldots, \lambda^{xy}_{2n}$.

\begin{Def}[Causal Structure] \label{defcausal} \normalfont 
\begin{leftbar}
The points~$x, y \in M$ are said to be
\[ \left\{ \begin{array}{cll}
{\mbox{\em{spacelike separated}}} &\quad& {\text{if all the~$\lambda^{xy}_j$ have the same absolute value}} \\[0.3em]
{\mbox{\em{timelike separated}}} && {\text{if the~$\lambda^{xy}_j$ are all real and do not all 
have }} \\
&& {\text{the same absolute value}} \\[0.3em]
{\mbox{\em{lightlike separated}}} && {\text{otherwise}}\:. \end{array}  \right. \]
\end{leftbar}
\end{Def} \noindent
This ``spectral definition'' of causality indeed gives back the causal structure
of Min\-kowski space or a Lorentzian manifold in the corresponding limiting cases
(for more details see Section~\ref{secapplication} below).
At this stage, one sees at least that our definition of the causal structure
is compatible with the Lagrangian in the
following sense. Suppose that two points~$x, y \in \F$ are spacelike separated.
Then the eigenvalues~$\lambda^{xy}_i$ all have the same absolute value,
implying that the Lagrangian vanishes.
Working out the corresponding Euler-Lagrange equations (for details 
see~\cite{jet}), one finds that
pairs of points with spacelike separation again drop out.
This can be seen in analogy to the usual notion of causality where
points with spacelike separation cannot influence each other.
In this sense, the {\em{principle of causality}} is built into the theory of causal fermion systems.

\subsection{Local Gauge Principle}
The fact that spacetime points of a causal fermion system are operators in~$\F$
gives rise to additional structures. In particular, there is an inherent notion of
spinors and wave functions, as we now explain.

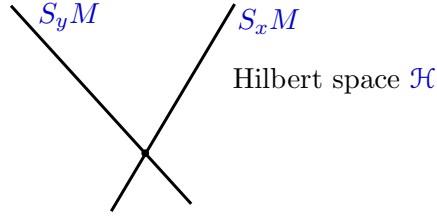
\begin{figure}% spinspace.svg
% \usepackage[usenames,dvipsnames]{pstricks}
% \usepackage{epsfig}
% \usepackage{pst-grad} % For gradients
% \usepackage{pst-plot} % For axes
% \usepackage[space]{grffile} % For spaces in paths
% \usepackage{etoolbox} % For spaces in paths
% \makeatletter % For spaces in paths
% \patchcmd\Gread@eps{\@inputcheck#1 }{\@inputcheck"#1"\relax}{}{}
% \makeatother
% \psscalebox{1.0 1.0} % Change this value to rescale the drawing.
{
\begin{pspicture}(-2,-1.3880072)(7.342354,1.3880072)
\rput[bl](2.962354,0.1906667){$\text{Hilbert space}~\Cb{\H}$}
\pscircle[linecolor=black, linewidth=0.02, fillstyle=solid,fillcolor=black, dimen=outer](1.7992429,-0.6066666){0.05111111}
\psline[linecolor=black, linewidth=0.04](0.014798415,1.3555557)(2.410354,-1.28)
\psline[linecolor=black, linewidth=0.04](2.9881318,1.3777778)(1.3481318,-1.3777777)
\rput[bl](3.0245762,0.98622227){$\Cb{S_xM}$}
\rput[bl](0.3756873,1.0217779){$\Cb{S_yM}$}
\end{pspicture}
}
\caption{The spin spaces}
\label{figspinspace}
\end{figure}

\begin{Def}[Spin Spaces] \normalfont \label{defspinspace}
\begin{leftbar}
	For every~$x \in M$ we define the {\em{spin space}}~$S_x$ by~$S_x = x(\H)$; it is a subspace of~$\H$ of dimension at most~$2n$.
On~$S_x$ we introduce an inner product $\Sl .|. \Sr_x$ by
\beq \label{sspintro}
\Sl . | . \Sr_x \::\: S_x \times S_x \rightarrow \C \:,\qquad
\Sl u | v \Sr_x = -\la u | x v \ra_\H \:,
\eeq
referred to as the {\em{spin scalar product}}. 
\end{leftbar}
\end{Def} \noindent
Since~$x$ has at most~$n$ positive
and at most~$n$ negative eigenvalues, the spin scalar product is an {\em{indefinite}} inner product
of signature~$(\p_x, \q_x)$ with~$\p_x, \q_x \leq n$
(for textbooks on indefinite inner product spaces see~\cite{bognar, GLR}).
In this way, to every spacetime point~$x \in M$ we associate
a corresponding indefinite inner product space~$(S_x, \Sl .|. \Sr_x)$.
If the signature of the spin spaces is constant in spacetime, we thus
obtain the structure of a topological vector bundle
(for more details in this direction see~\cite{topology}).
However, in contrast to a vector bundle, all the spin spaces are
subspaces of the same Hilbert space~$\H$; see Figure~\ref{figspinspace}.

\begin{figure} % wavefunction.svg
% \usepackage[usenames,dvipsnames]{pstricks}
% \usepackage{epsfig}
% \usepackage{pst-grad} % For gradients
% \usepackage{pst-plot} % For axes
% \usepackage[space]{grffile} % For spaces in paths
% \usepackage{etoolbox} % For spaces in paths
% \makeatletter % For spaces in paths
% \patchcmd\Gread@eps{\@inputcheck#1 }{\@inputcheck"#1"\relax}{}{}
% \makeatother
% \psscalebox{1.0 1.0} % Change this value to rescale the drawing.
{
\begin{pspicture}(0,-1.3880072)(3.7745762,1.3880072)
\definecolor{colour0}{rgb}{0.8,0.2,0.0}
\definecolor{colour1}{rgb}{0.8,0.0,0.0}
\pscircle[linecolor=black, linewidth=0.02, fillstyle=solid,fillcolor=black, dimen=outer](1.7992429,-0.6066666){0.05111111}
\psline[linecolor=black, linewidth=0.04](0.014798415,1.3555557)(2.410354,-1.28)
\psline[linecolor=black, linewidth=0.04](2.9881318,1.3777778)(1.3481318,-1.3777777)
\rput[bl](3.0245762,0.98622227){$\Cb{S_xM}$}
\rput[bl](0.3756873,1.0217779){$\Cb{S_yM}$}
\psline[linecolor=colour0, linewidth=0.04, arrowsize=0.05291667cm 2.0,arrowlength=1.4,arrowinset=0.0]{->}(1.8059095,-0.60888886)(1.3392428,1.1422223)
\psline[linecolor=colour1, linewidth=0.02, linestyle=dashed, dash=0.17638889cm 0.10583334cm](1.3481318,1.12)(2.4281318,0.4666667)
\psline[linecolor=colour1, linewidth=0.02, linestyle=dashed, dash=0.17638889cm 0.10583334cm](1.3436873,1.1244445)(0.7836873,0.53333336)
\psarc[linecolor=colour1, linewidth=0.02, dimen=outer](2.4392428,0.45777783){0.2911111}{150.0}{240.0}
\psarc[linecolor=colour1, linewidth=0.02, dimen=outer](0.7725762,0.49777782){0.2911111}{-40.0}{50.0}
\pscircle[linecolor=colour1, linewidth=0.02, dimen=outer](0.9436873,0.5155556){0.02}
\pscircle[linecolor=colour1, linewidth=0.02, dimen=outer](2.281465,0.42222226){0.02}
\rput[bl](1.4690206,1.1684445){\textcolor{colour0}{$u$}}
\rput[bl](2.2734652,-0.2804444){\textcolor{colour0}{$\psi^u(x)$}}
\rput[bl](0.22013175,-0.3693333){\textcolor{colour0}{$\psi^u(y)$}}
\end{pspicture}
}
\caption{The physical wave function}
\label{figwavefunction}
\end{figure}
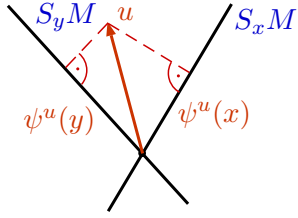

The vectors in~$\H$ can be represented as wave functions in spacetime:
\begin{Def}[Physical Wave Function] \label{defpwf} \normalfont 
\begin{leftbar}
For a vector~$u \in \H$ one introduces the corresponding
{\em{physical wave function}} $\psi^u$ as
\[ \psi^u \::\: M \rightarrow \H\:,\qquad \psi^u(x) = \pi_x u \in S_x \:, \]
where~$\pi_x : \H \rightarrow S_x$ denotes the orthogonal projection on the subspace~$S_x \subset \H$.
\end{leftbar}
\end{Def} \noindent
This definition is illustrated in Figure~\ref{figwavefunction}.

A {\em{local gauge principle}} becomes apparent once we choose
basis representations of the spin spaces and write the wave functions in components.
Denoting the signature of~$(S_x, \Sl .|. \Sr_x)$ by~$(\p_x,\q_x)$, we choose
a pseudo-orthonormal basis~$(\mathfrak{e}_\alpha(x))_{\alpha=1,\ldots, \p_x+\q_x}$ of~$S_x$, i.e.
\[ \Sl \mathfrak{e}_\alpha(x) | \mathfrak{e}_\beta(x) \Sr_x = s_\alpha\: \delta^\alpha_\beta \]
with~$s_1=\ldots=s_{\p_x}=1$ and~$s_{\p_x+1}=\ldots=s_{\p_x+\q_x}=-1$.
Then a physical wave function~$\psi^u$ can be represented as
\[ \psi^u(x) = \sum_{\alpha=1}^{\p_x+\q_x} \psi^\alpha(x)\: \mathfrak{e}_\alpha(x) \]
with component functions~$\psi(x)^1, \ldots, \psi(x)^{\p_x+\q_x}$.
The freedom in choosing the basis~$(\mathfrak{e}_\alpha)$ is described by the
group~$\U(\p_x, \q_x)$ of unitary transformations with respect to an inner product of signature~$(\p_x,\q_x)$.
This gives rise to the transformations
\beq \label{gauge1}
\mathfrak{e}_\alpha(x) \rightarrow \sum_{\beta=1}^{\p_x+\q_x} U^{-1}(x)^\beta_\alpha\;
\mathfrak{e}_\beta(x) \qquad \text{and} \qquad
\psi^\alpha(x) \rightarrow  \sum_{\beta=1}^{\p_x+\q_x} U(x)^\alpha_\beta\: \psi^\beta(x)
\eeq
with $U \in \U(\p_x, \q_x)$.
As the basis~$(\mathfrak{e}_\alpha)$ can be chosen independently at each space-time point,
one obtains {\em{local gauge transformations}} of the wave functions,
where the gauge group is determined to be the isometry group of the spin scalar product.

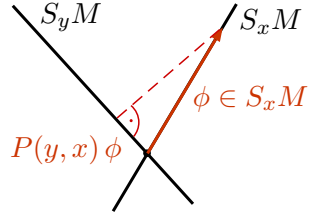
\begin{figure} % kernel.svg
% \usepackage[usenames,dvipsnames]{pstricks}
% \usepackage{epsfig}
% \usepackage{pst-grad} % For gradients
% \usepackage{pst-plot} % For axes
% \usepackage[space]{grffile} % For spaces in paths
% \usepackage{etoolbox} % For spaces in paths
% \makeatletter % For spaces in paths
% \patchcmd\Gread@eps{\@inputcheck#1 }{\@inputcheck"#1"\relax}{}{}
% \makeatother
% \psscalebox{1.0 1.0} % Change this value to rescale the drawing.
{
\begin{pspicture}(0,-1.3880072)(3.8166666,1.3880072)
\definecolor{colour0}{rgb}{0.8,0.2,0.0}
\definecolor{colour1}{rgb}{0.8,0.0,0.0}
\pscircle[linecolor=black, linewidth=0.02, fillstyle=solid,fillcolor=black, dimen=outer](1.8191111,-0.6066666){0.05111111}
\psline[linecolor=black, linewidth=0.04](0.034666665,1.3555557)(2.4302223,-1.28)
\psline[linecolor=black, linewidth=0.04](3.008,1.3777778)(1.368,-1.3777777)
\rput[bl](3.0444446,0.98622227){$S_xM$}
\rput[bl](0.39555556,1.0217779){$S_yM$}
\psline[linecolor=colour0, linewidth=0.04, arrowsize=0.05291667cm 2.0,arrowlength=1.4,arrowinset=0.0]{->}(1.8257778,-0.60888886)(2.8257778,1.0666667)
\psline[linecolor=colour1, linewidth=0.02, linestyle=dashed, dash=0.17638889cm 0.10583334cm](2.7857778,1.0400001)(1.4213333,-0.13777773)
\psarc[linecolor=colour1, linewidth=0.02, dimen=outer](1.4191111,-0.19555551){0.2911111}{-40.0}{50.0}
\pscircle[linecolor=colour1, linewidth=0.02, dimen=outer](1.608,-0.17333329){0.02}
\rput[bl](2.4266667,-0.04488885){\textcolor{colour0}{$\phi \in S_xM$}}
\rput[bl](0.0,-0.756){\textcolor{colour0}{$P(y,x)\, \phi$}}
\end{pspicture}
}
\caption{The kernel of the fermionic projector}
\label{figkernel}
\end{figure}

The causal action is {\em{gauge invariant}} in the sense that it does not depend on the choice of spinor
bases. This connection becomes clearer if the Lagrangian is expressed in terms of the physical
wave functions. This can be accomplished as follows.
\begin{Def}[Kernel of the Fermionic Projector] \normalfont 
\begin{leftbar}
For any~$x, y \in M$ we define the
{\em{kernel of the fermionic projector}}~$P(x,y)$ by
\beq \label{Pxydefintro}
P(x,y) = \pi_x \,y|_{S_y} \::\: S_y \rightarrow S_x
\eeq
\end{leftbar}
\end{Def} \noindent
This definition is illustrated in Figure~\ref{figkernel}.
We remark that this definition harmonizes with the definition of the
spin scalar product~\eqref{sspintro} in the sense that the kernel of the fermionic projector
is {\em{symmetric}} with respect to the spin scalar product,
\begin{align*}
\Sl u \,|\, P(x,y) \,v \Sr_x &= - \la u \,|\, x\, P(x,y) \,v \ra_\H = - \la u \,|\, x y \,v \ra_\H \\
&= -\la \pi_y \,x\, u \,|\, y \,v \ra_\H = \Sl P(y,x)\, u \,|\,  v \Sr_y
\end{align*}
(where~$u \in S_x$ and~$v \in S_y$).
Taking the trace of~\eqref{Pxydefintro} in the case~$x=y$, one
finds that~$\tr(x) = \Tr_{S_x}(P(x,x))$
(where~$\tr$ and~$\Tr_{S_x}$ are the traces on~$\H$ and the spin space, respectively),
making it possible to express
the integrand of the trace constraint~\eqref{trconstraint} in terms of the kernel of the fermionic projector.
In order to also express the eigenvalues of the operator~$xy$ in terms of the kernel of the fermionic projector,
we introduce the {\em{closed chain}}~$A_{xy}$ as the product
\beq \label{Axydef}
A_{xy} = P(x,y)\, P(y,x) \::\: S_x \rightarrow S_x\:.
\eeq
Computing powers of the closed chain, one obtains
\[ A_{xy} = (\pi_x y)(\pi_y x)|_{S_x} = \pi_x\, yx|_{S_x} \:,\qquad
(A_{xy})^p = \pi_x\, (yx)^p|_{S_x} \:. \]
Taking the trace, one sees in particular that
\[ %\label{trid}
\Tr_{S_x}(A_{xy}^p) = \tr \big((yx)^p \big) = \tr \big((xy)^p \big) \]
(where the last identity simply is the invariance of the trace under cyclic permutations).
As a consequence\footnote{More precisely, since all our operators have finite rank, there is a finite-dimensional subspace~$I$
of~$\H$ such that~$xy$ maps~$I$ to itself and vanishes on the orthogonal complement of~$I$.
Then the non-trivial eigenvalues of the operator product~$xy$ are given
as the zeros of the characteristic polynomial of the restriction~$xy|_I : I \rightarrow I$.
The coefficients of this characteristic polynomial (like the trace, the determinant, etc.)
are symmetric polynomials in the eigenvalues and can therefore be expressed in terms of traces of
powers of~$A_{xy}$.}, the eigenvalues of the closed chain coincide with the non-trivial
eigenvalues~$\lambda^{xy}_1, \ldots, \lambda^{xy}_{2n}$ of the operator product~$xy$.
This makes it possible to express both the Lagrangian~\eqref{Ldef}
and the integrand of the boundedness constraint~\eqref{Tdef} in terms of~$A_{xy}$.
The main advantage of working with the kernel of the fermionic projector is that the closed chain~\eqref{Axydef}
is a linear operator on a vector space of dimension at most~$2n$, making it possible
to compute the~$\lambda^{xy}_1, \ldots, \lambda^{xy}_{2n}$ as the eigenvalues of 
a finite matrix.

The kernel of the fermionic projector can be expressed in terms of the physical
wave functions as follows. Choosing an orthonormal basis~$(e_i)$ of~$\H$ and using the completeness relation
as well as~\eqref{sspintro}, one obtains
for any~$\phi \in S_y$
\[ P(x,y) \, \phi= \pi_x y|_{S_y} \,\phi = \sum_i \pi_x e_i \: \la e_i | y\,\phi \ra_\H
= -\sum_i \psi^{e_i}(x) \: \Sl \psi^{e_i}(y) \,|\, \phi \Sr_y  \:, \]
showing that~$P(x,y)$ is indeed composed of all the physical wave functions, i.e.\
in a bra/ket notation
\beq \label{Prep}
P(x,y) = -\sum_i |\psi^{e_i}(x) \Sr\: \Sl \psi^{e_i}(y)|  \:.
\eeq
Finally, choosing again bases~$(\mathfrak{e}_\alpha(x))_{\alpha=1,\ldots, \p_x+\q_x}$ of the
spin spaces, the kernel~$P(x,y)$ is expressed by a~$(\p_x+\q_x) \times (\p_y + \q_y)$-matrix.
According to~\eqref{gauge1}, this matrix behaves under gauge transformations as
\[ P(x,y)^\alpha_\beta \rightarrow \sum_{\gamma=1}^{\p_x+\q_x} \sum_{\delta=1}^{\p_y+\q_y} 
U(x)^\alpha_\gamma \:P(x,y)^\gamma_\delta \: \big(U(y)^* \big)^\delta_\beta \:, \]
where the star denotes the adjoint with respect to the spin scalar product.
Since~$U(y) \in U(\p_x, \q_x)$ is unitary with respect to the spin scalar product, 
the gauge transformation at~$y$ drops out when forming the closed chain, i.e.
\[ (A_{xy})^\alpha_\beta \rightarrow \sum_{\gamma, \delta=1}^{\p_x+\q_x}
U(x)^\alpha_\gamma \:(A_{xy})^\gamma_\delta \: \big(U(x)^* \big)^\delta_\beta \:. \]
Since~$U(x) \in U(\p_x, \q_x)$ is unitary, the eigenvalues of the closed chain
do not depend on the choice of the gauge.

This explains in particular why the Lagrangian is invariant under local gauge
transformations of the physical wave functions.
Such computations were helpful for formulating the causal action principle
(for details see~\cite[Chapter~3]{pfp}).

\subsection{Fermionic Building Blocks}
In the above formulas, the physical wave functions play a dominant role.
Indeed, according to~\eqref{Prep}, the ensemble of all these wave functions
determines the kernel of the fermionic projector, which, forming the closed
chain and computing its eigenvalues, gives rise to all the quantities needed in
the causal action principle. In this way, the causal variational principle
can be formulated directly in terms of the ensemble of all physical wave functions.
%\mpar{Besser hervorheben? The collective behaviour of all physical wave functions then gives us the world as we know it as emergent structures. }%
Minimizing the causal action amounts to finding an ``optimal'' configuration of the physical
wave functions.
In other words, the causal action principle can be understood as a
variational principle which determines the collective behavior of all physical wave functions.

As will be worked out in detail in Section~\ref{secapplication} below,
in concrete examples the physical wave functions go over to solutions
of the Dirac equation. More specifically, describing the Minkowski vacuum
as a causal fermion system (see Section~\ref{secmink}), the ensemble of all physical wave functions
correspond to all the negative-frequency solutions of the Dirac equation.
In this way, Dirac's original concept of the Dirac sea is realized.
The fact that Dirac wave functions describe fermionic particles is the motivation
for the name ``causal {\em{fermion}} system.''

\subsection{Microscopic Spacetime Structure}
In the theory of causal fermion systems, spacetime defined as the support
of the universal measure~$\rho$ (see Definition~\ref{defst}) does not need to be a
differentiable manifold. Instead, it could be discrete on a microscopic scale
or could have another nontrivial microstructure.
Exactly as explained above for the causal structure, also the microscopic
structure of spacetime is not given a-priori, but it is determined dynamically by the
causal action principle. The analysis of simple model examples
reveals that minimizing measures of the causal action principles are typically
discrete (for details see~\cite{support, sphere} or the survey in~\cite[Section~3]{dice2018}).
Although it is an open problem whether these discreteness results also hold
for general causal fermion systems, these results suggest
that the concept of smooth spacetime structures
should be modified on small scales, typically thought of as the Planck scale.
The theory of causal fermion systems provides a mathematical setting in which
such generalized spacetimes can be described and analyzed.

\section{Modelling a Lorentzian Spacetime by a Causal Fermion System} \label{secapplication}
\subsection{General Construction in Curved Spacetimes} \label{seccurv}
We return to the setting of the Dirac equation in curved spacetime in Section~\ref{secQMcurved}.
We now explain how to describe this spacetime by a causal fermion system.
We denote the Hilbert space of solutions of the Dirac equation with the scalar product~\eqref{print}
by~$(\H_m, (.|.)_m)$ (more precisely, we take the completion of all smooth solutions with
spatially compact support). 
Next, we choose a closed subspace~$\H \subset \H_m$
of the solution space of the Dirac equation.
The induced scalar product on~$\H$ is denoted by~$\la .|. \ra_\H$.
There is the technical difficulty that the wave functions in~$\H$ are in general not continuous,
making it impossible to evaluate them pointwise.
For this reason, we need to introduce an {\em{ultraviolet regularization}}, described mathematically by a linear
\[ \text{\em{regularization operator}} \qquad {\mathfrak{R}} \::\: \H \rightarrow C^0(\scrM, S\scrM) \:. \]
We postpone the discussion of the physical significance of the regularization operator to
Section~\ref{secregop}. Mathematically, the simplest method to obtain
a regularization operator is by taking the convolution with a smooth, compactly supported function
on a Cauchy surface or in spacetime (for details see~\cite[Section~4]{finite}
or~\cite[Section~\S1.1.2]{cfs}).

Given~${\mathfrak{R}}$, for any space-time point~$p \in \scrM$ we consider the bilinear form
\[ b_p \::\: \H \times \H \rightarrow \C\:,\quad
b_p(\psi, \phi) = -\Sl ({\mathfrak{R}}\psi)(p) | ({\mathfrak{R}} \phi)(p) \Sr_p \:. \]
This bilinear form is well-defined and bounded because~${\mathfrak{R}}$ 
maps to the continuous wave functions and because evaluation at~$p$ gives a linear operator of finite rank.
Thus for any~$\phi \in \H$, the anti-linear form~$b_p(.,\phi) : \H \rightarrow \C$
is continuous. By the Fr{\'e}chet-Riesz theorem,
there is a unique~$\chi \in \H$
such that~$b_p(\psi,\phi) = \la \psi | \chi \ra_\H$ for all~$\psi \in \H$.
The mapping~$\phi \mapsto \chi$ is linear and bounded, giving rise to the following
linear operator:
\begin{Def}[Local Correlation Operator] \normalfont 
\begin{leftbar}
For any~$p \in \scrM$, the {\em{local correlation operator}}~$F(p)$ on~$\H$
is defined by the relation
\beq \label{Fepsdef}
(\psi \,|\, F(p)\, \phi) = -\Sl ({\mathfrak{R}}\psi)(p) | 
({\mathfrak{R}} \phi)(p) \Sr_p \qquad \text{for all~$\psi, \phi
\in \H$} \:.
\eeq
\end{leftbar}
\end{Def} \noindent
Taking into account that the inner product on the Dirac spinors at~$p$ has signature~$(2,2)$,
the local correlation operator~$F(p)$ is a symmetric operator on~$\H$
of rank at most four, which (counting multiplicities) has at most two positive and at most two negative eigenvalues.
Varying the space-time point, we obtain a mapping
\[ F \::\: \scrM \rightarrow \F \subset \Lin(\H)\:, \]
where~$\F$ denotes all symmetric operators of
rank at most four with at most two positive and at most two negative eigenvalues.
Finally, we introduce the
\beq \label{pushforward}
\text{\em{universal measure}} \qquad d\rho := F_* \,d\mu_\scrM
\eeq
as the push-forward of the volume measure on~$\scrM$ under the mapping~$F$
(thus~$\rho(\Omega) := \mu_\scrM(F^{-1}(\Omega))$).
We thus obtain a causal fermion system~$(\H, \F, \rho)$ of spin dimension two.

We close with a few comments
on the underlying physical picture.
The vectors in the subspace~$\H \subset \H_m$ have the interpretation
as those Dirac wave functions which are realized in the physical system under
consideration.
If we describe for example a system of one electron,
then the wave function of the electron is contained in~$\H$. Moreover, $\H$ includes all the wave functions
which form the so-called Dirac sea (for an explanation of this point
see for example~\cite{srev}). 

According to~\eqref{Fepsdef}, 
the local correlation operator~$F(p)$ describes 
densities and correlations of the physical
wave functions at the space-time point~$p$.
Working exclusively with the local correlation operators and the
corresponding push-forward measure~$\rho$ means in particular
that the geometric structures are encoded in and must be retrieved from the physical wave functions.
Since the physical wave functions describe the distribution of
matter in space-time, one can summarize this concept
by saying that {\em{matter encodes geometry}}.
Going one step further, one can also say that matter and geometry form an inseparable unity.
%and therefore cannot be treated independently.

%\mpar{Sage hier mehr: Geometrie und Materie bilden eine Einheit? Claudio:
%I'm not sure i m 100\% happy with this statement / or there could actually be more to it. the point is : taking all the modified connection (which contains all bosonic fields) the curvature of this connection determines ALL forces) i.e. the geometry encodes all forces between particels but the geometry of the modiifed conenction... i think at that point lies some deeper conceptional insight that one might want to explore }%

\subsection{Physical Significance of the Regularization Operator} \label{secregop}
The regularization operator requires a detailed explanation.
We first convey the underlying physical picture.
The regularization operators should leave the wave functions unchanged on macroscopic
scales (i.e.\ scales much larger than the Planck length).
Thus on macroscopic length scales, the Dirac equation still holds,
giving agreement with the common physical description.
However, on a microscopic scale~$\varepsilon$, which can be thought of as the Planck scale,
the regularization may change the wave functions completely.
As a consequence, also the universal measure~$\rho$ in~\eqref{pushforward} is changed,
which means that the microscopic structure of spacetime is modified.
Therefore, in contrast to the renormalization program in Quantum Field Theory,
in the theory of causal fermion systems the regularization is not just a technical tool,
but it realizes our concept that we want to allow for a nontrivial microstructure
of spacetime. With this in mind, we always consider the regularized quantities as
those having mathematical and physical significance. Different choices of regularization operators
realize different microscopic spacetime structures.

This concept immediately raises the question how the ``physical regularization'' should look like.
Generally speaking, the regularized spacetime should look like Lorentzian spacetime down to distances
of the scale~$\varepsilon$. For distances smaller than~$\varepsilon$, the structure of
space-time may be completely different, in a way which cannot be 
guessed or extrapolated from the structures of Minkowski space.
Since experiments on the length scale~$\varepsilon$ seem out of reach, it is completely
unknown what the microscopic structure of space-time is.
Within the theory of causal fermion systems, the above question could be answered in principle
by minimizing the causal action over all possible regularization operators.
However, this approach turns out to be very difficult and at present is out of reach
(for a first step in this direction see~\cite{reg}).
In view of these difficulties, the only available method is the so-called
{\em{method of variable regularization}}:
Instead of trying to determine the microstructure experimentally or with mathematical analysis,
the strategy is to a-priori include all conceivable regularizations and, with hindsight, to eliminate those
which are in conflict with well-established physical facts. The remaining regularizations which comply with all
experimental constraints should be treated as equally admissible, because at present
there is no criterion to distinguish between different choices or to favor one regularization over the others.

For the method of variable regularization to be sensible and to retain the predictive power of the theory,
the detailed form of the microstructure must have no influence
on the effective physical equations which are valid on the energy scales accessible to experiments.
More precisely, the picture is that the general structure of the effective physical equations should be independent
of the microstructure of spacetime. Values of mass ratios or coupling constants, however, may well
depend on the microstructure (a typical example is the gravitational constant, which is
closely tied to the Planck length).
In more general terms, the unknown microstructure
of spacetime should enter the effective physical equations only by a finite (hopefully
small) number of free parameters, which can then be taken as empirical free parameters
of the effective macroscopic theory. In~\cite{cfs} it was shown that these conditions are indeed
satisfied.

\subsection{Concrete Example: the Minkowski Vacuum} \label{secmink}
We now make the construction of Section~\ref{seccurv} more explicit by working out
the example of the Minkowski vacuum with the simplest possible regularization.
We proceed in the following steps: \vspace*{2.0mm}
\bitem \setlength\itemsep{1mm}
\itemD \textbf{Choosing the Hilbert space of all negative-frequency solutions} \\
Our starting point are the plane-wave solutions of the Dirac equation
in Minkowski space~\eqref{Dirac1}, which we write as
\[ \psi_{\vec{p}a\pm}(x)=\frac{1}{\sqrt{(2\pi)^3}}\mathrm{e}^{\mp\mathrm{i}\omega t+\mathrm{i}\vec{p}\cdot\vec{x}}\chi_{\vec{p}a\pm}\qquad\text{with}\qquad\omega=\omega(\,\vec{p}\,):=\sqrt{|\,\vec{p}\,|^2+m^2} \:. \]
Here the spinor $\chi_{\vec{p}a\pm}$ solves the algebraic equation
\[ %\begin{equation}\label{eq:spinor_equation}
(\gamma^kp_k-m\mathbbm{1})\chi_{\vec{p}a\pm}=0 \:, \]
where $(p_k)=(\omega,\vec{p}\,)$ denotes the four-momentum. Negative-frequency wave packets of the form
\beq \label{packet}
\psi_f(x) := \int_{\R^3} \psi_{\vec{p}a-}(x)\: f(\vec{p}\,)\: d^3p \qquad \text{with} \qquad f \in C^\infty_0(\R^3, \C)
\eeq
span a subspace of~$\H_m$. We choose the Hilbert space~$\H$ of the causal fermion system
as the closure of this subspace. This choice realize the concept of the Dirac sea vacuum.
\itemD \textbf{Constructing the local correlation operators} \\
The simplest method to choose regularization operators consists in inserting a convergence-generating factor~$e^{-\varepsilon \omega}$ into the wave packet~\eqref{packet}, i.e.
\[ \big({\mathfrak{R}} \psi_f \big)(x) := \int_{\R^3} e^{-\varepsilon \omega} \psi_{\vec{p}a-}(x)\: f(\vec{p}\,)\: d^3p \:. \]
Now we can define the local correlations operators by~\eqref{Fepsdef}
and construct the universal measure according to~\eqref{pushforward}. We thus obtain
a causal fermion system~$(\H, \F, \rho)$.
\eitem
In this example, one can compute the objects of the causal fermion system explicitly
(for details see~\cite[Section~1.2]{cfs}). One finds that in the limit~$\varepsilon \searrow 0$,
the inherent structures of the causal fermion system go over to the usual 
objects and relations in Minkowski space. More specifically, mapping a point~$p \in \scrM$
to the corresponding local correlation operator~$F(p)$ gives a one-to-one correspondence
between Minkowski space~$\scrM$ and the spacetime~$M:= \supp \rho$ of the causal fermion system.
Moreover, the causal structure of Definition~\ref{defcausal}
gives back the causal structure of Minkowski space, and the spin space~$S_x$ of Definition~\ref{defspinspace}
can be identified with the space of Dirac spinors~$S_p\scrM$.
Under these identifications, the physical wave functions of Definition~\ref{defpwf} agree with the
regularized Dirac wave functions of negative frequency.

\section{Results of the Theory and Further Reading}
Let us explain in which sense and to which extent the goal of unifying Quantum Field Theory
and General Relativity has been achieved. Causal fermion systems provide a mathematically
consistent theory which gives General Relativity and Quantum Theory as limiting cases.
The causal action principle has well-defined minimizers in the
case of a finite-dimensional Hilbert space
and finite total volume (see~\cite{continuum}; more general cases are presently under investigation).
The reason why the inconsistencies
of Quantum Field Theory and General Relativity as described in Section~\ref{secinconsist}
as well as the divergences of Quantum Field Theory disappear is that we modified the structure
of spacetime on the Planck scale. In more technical terms, in a causal fermion system one works
with the regularized objects. Thus we consider the regularized objects as the fundamental physical
objects. This concept could be implemented coherently because the causal action principle is
formulated purely in terms of these regularized objects.

Causal fermion systems are a unified theory in the sense that spacetime and all objects therein are
described by a single object: the universal measure. The causal action principle singles out those
measures which describe physically admissible spacetimes. The Euler-Lagrange equations corresponding
to the causal action principle describe the spacetime dynamics.

Clearly, in this short review we could only cover certain aspects of the theory from a particular perspective.
Therefore, in order to help the interested reader to get a more complete picture,
we now outline a few other directions and give references for further study.
For other review articles with a somewhat different focus we refer to~\cite{srev, rrev, dice2014}.
\begin{itemize}[leftmargin=2em]
\item[(a)] A causal fermion system also provides {\em{topological}} (topological spinor bundle)
and {\em{geometric objects}} (parallel transport and curvature). We refer the interested reader
to~\cite{lqg, topology} or the introduction~\cite{nrstg}.
\item[(b)] The limiting case~$\varepsilon \searrow 0$, when the ultraviolet regularization is removed,
is worked out in detail in~\cite{cfs}. In this limiting case, the so-called {\em{continuum limit}},
the causal action principle gives rise to
the interactions of the standard model and gravity, on the level of classical bosonic fields
interacting with a second-quantized fermionic field.
\item[(c)] An important concept for more recent developments are {\em{surface layer integrals}},
which generalize surface integrals to the setting of causal fermion systems.
Symmetries of causal fermion systems give rise to conservation laws which can be expressed
in terms of surface layer integrals~\cite{noether}.
\item[(d)] Another concept which has turned out to be fruitful for the analysis of the causal
action principle are {\em{linearized solutions}}~\cite{jet}. Similar to linearized gravitational waves,
linearized solutions can be understood as linear perturbations of the measure~$\rho$ which
preserve the Euler-Lagrange equations of the causal action principle.
As shown in~\cite{jet, osi}, linearized solutions come with corresponding conserved
surface layer integrals, in particular the {\em{symplectic form}} and the {\em{surface layer inner product}}.
\item[(e)] Generally speaking, the conservation laws for surface layer integrals give rise to
objects in space which evolve dynamically in time.
This concept was worked out for linearized solutions in~\cite{linhyp}, where it is proven
under general assumptions that the {\em{Cauchy problem}} for linearized solutions is well-posed
and that the solutions propagate with finite speed.
\item[(f)] A first connection to {\em{Quantum Field Theory}} has been made in~\cite{qft},
however based on the classical field equations obtained in the continuum limit.
Deriving Quantum Field Theory as a limiting case of causal fermion systems
without referring to the continuum limit is a major objective of present research:
The perturbation theory for the universal measure is worked out in~\cite{perturb}.
For interacting bosonic fields, the constructions in~\cite{fockbosonic}
give rise to a description of the dynamics in terms of a unitary time evolution on bosonic Fock spaces.
The generalization of these constructions to include fermionic fields is currently under investigation~\cite{fockfermionic}.
\end{itemize}

%eventuell noch erw\"ahnen:
%\begin{itemize}
%\item cosomological applications: \cite{dgc}, \cite{reghadamard}
%\end{itemize}

\Thanks{{{\em{Acknowledgments:}} We would like to thank the participants of the
conference ``Progress and visions in quantum theory in view of gravity'' held in Leipzig in October 2018
for fruitful and inspiring discussions. We would like to thank Jos{\'e} M.\ Isidro, Christoph Langer,
Claudio Paganini, Julian Seipel and the referee for helpful comments on the manuscript.
M.~J.\ gratefully acknowledges support by the ``Studienstiftung des deutschen Volkes'' and the ``Hanns-Seidel-Stiftung.''

%\newpage
%\bibliographystyle{amsplain}
%\bibliography{../../aarbeit/felix}
\providecommand{\bysame}{\leavevmode\hbox to3em{\hrulefill}\thinspace}
\providecommand{\MR}{\relax\ifhmode\unskip\space\fi MR }
\providecommand{\MRhref}[2]{%
  \href{http://www.ams.org/mathscinet-getitem?mr=#1}{#2}
}
\providecommand{\href}[2]{#2}

\end{document}